\documentclass[aps,twocolumn,prl,preprintnumbers,amsmath,amssymb,superscriptaddress]{revtex4}

\usepackage{graphicx}
\usepackage{bm}

\begin{document}

\title{On the possibility of fast vortices in the Cuprates: A vortex plasma model analysis of the THz conductivity and 
diamagnetism in La$_{2-x}$Sr$_{x}$CuO$_4$}

\author{L.S. Bilbro}
\affiliation{The Institute for Quantum Matter, Department of Physics and Astronomy, The Johns Hopkins University, Baltimore, MD 21218, USA}

\author{R.Vald\'es Aguilar}
\affiliation{The Institute for Quantum Matter, Department of Physics and Astronomy, The Johns Hopkins University, Baltimore, MD 21218, USA}

\author{G. Logvenov}
\affiliation{Brookhaven National Laboratory, Upton, NY 11973, USA}

\author{I. Bozovic}
\affiliation{Brookhaven National Laboratory, Upton, NY 11973, USA}

\author{N.P. Armitage}
\affiliation{The Institute for Quantum Matter, Department of Physics and Astronomy, The Johns Hopkins University, Baltimore, MD 21218, USA}

\date{\today}

\begin{abstract}

We present measurements of the fluctuation superconductivity in an underdoped thin film of La$_{1.905}$Sr$_{0.095}$CuO$_4$  using time-domain THz spectroscopy.  We compare our results with measurements of diamagnetism in a similarly doped crystal of La$_{2-x}$Sr$_x$CuO$_4$.  We show through a vortex-plasma model that if the fluctuation diamagnetism solely originates in vortices, then they must necessarily exhibit an anomalously large vortex diffusion constant, which is more than two orders of magnitude larger than the Bardeen-Stephen estimate.  This points to either the extremely unusual properties of vortices in the under-doped $d$-wave cuprates or a contribution to the diamagnetic response that is not superconducting in origin.

\end{abstract}



\maketitle

Nearly 25 years after the demonstration of high-temperature superconductivity in the cuprate superconductors and more than 15 years since the discovery of the anomalous pseudogap in underdoped compounds, the microscopic physics of the superconducting phase and its relationship to the pseudogap remain hotly debated.  Due to their low superfluid densities, it is generally agreed that superconducting fluctuations will be large and prominent in these materials \cite{Emery95a}.  What is less agreed upon is the temperature range above $T_c$ in which superconducting correlations are truly significant and their contributions to the physics of the pseudogap.  Experimental probes such as photoemission, tunneling, NMR spin relaxation, heat capacity, the Nernst effect, and diamagnetic susceptibility have shown evidence for a gaplike structure reminiscent of $d$-wave superconductivity in the density of states implying a strong connection of the pseudogap to superconductivity and/or superconducting correlations at temperatures well above $T_c$ \cite{Timusk99a,Norman05a,Xu00a,Wang05a,Li10a}.  However, other mechanisms exist that can create such structures in the density of states \cite{Hlubina95a,Chakravarty01a}.

\begin{figure*}[t]
\includegraphics[width=2.1\columnwidth]{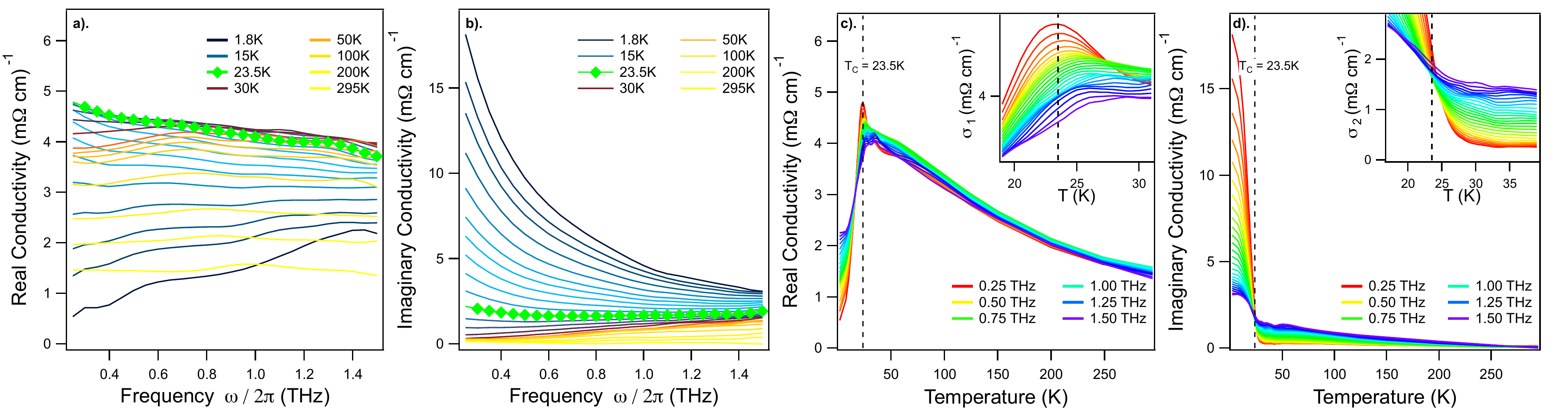}
\caption{(a) Real and (b) imaginary conductivities as a function of frequency at different temperatures of a $x=0.095$ $T_c = 23.5$ K LSCO film. (c) Real and (d) imaginary conductivities as a function of temperature at different frequencies.  In panels (a) and (b) the green curve denotes $T_c$.  In panels (c) and (d) the vertical lines represent $T_c$.  Insets to (c) and (d) show expanded views of the fluctuation region.}
\end{figure*}

Interestingly, perhaps the most essential probe of the electronic properties -- charge transport -- does not show an extended range of superconducting fluctuations in temperature or field. \cite{Corson99a,Miura02a,Ando95a}.  In La$_{2-x}$Sr$_x$CuO$_4$ the region of enhanced diamagnetism extends almost 100 K above $T_c$ \cite{Li10a} while the THz fluctuation conductivity has an extent limited to 10 - 20 K above $T_c$ \cite{Bilbro11a}.  This is surprising as one might expect a close correspondence between these quantities \cite{Halperin79a}.  Similarly, it has been argued from Nernst and diamagnetism measurements that $H_{c2}$ may be as high as 150 T \cite{Li10a}, while the resistive transition is essentially complete in optimally and underdoped LSCO by 45 T \cite{Miura02a,Ando95a}.   

In this Rapid Communication we present results of our detailed THz time-domain spectroscopy (TTDS) study of the fluctuation superconductivity in LSCO.  The THz fluctuation conductivity shows an onset approximately only 10 K above $T_c$, which contrasts strongly with measurements like diamagnetism in which the onset is approximately 100K above $T_c$.  We analyze our data in the context of a vortex plasma model and show, however that it is not the functional dependences of these data that are in strongest contrast, but their overall scales.  Conventional vortex dynamics would predict a much larger fluctuation conductivity given the size of diamagnetism.  We demonstrate that if the regime of enhanced diamagnetism originates in vortices, then the vortex diffusion constant $D$ must be anomalously large and in the range of 10-30 cm$^2/$sec above $T_c$.   This is more than two orders of magnitude larger than conventional benchmarks based on the Bardeen-Stephen model \cite{Stephen65a}.  It is then a well-posed theoretical challenge to explain a $D$ this large.  This points to either extremely unusual vortex properties in the underdoped $d$-wave cuprates or a contribution to the diamagnetic response that is not superconducting in origin.

We begin with the observation that the ratio $\chi_{2D}/\mu_0 G$ of the two-dimensional (2D) susceptibility over the conductance has units of length squared over time, i.e., diffusion \cite{Torron94a}.  One can show that in a diffusive vortex plasma this ratio gives a unique measure of the vortex diffusion constant\cite{Orenstein06a}.  Using the notation of Halperin and Nelson \cite{Halperin79a}, but in SI units, the 2D susceptibility and conductance of a conventional thin superconducting film at temperatures above a vortex unbinding transition are
\begin{eqnarray}
\chi_{2D} = - \frac{ c_2 \pi^2\mu_0 k_B T}{\phi_0^2} \xi^2
\label{Suscp}\\
G_S = \frac{1}{\phi_0^2 n_f \mu. }
\label{Cond}
\end{eqnarray}

 \noindent Here $\xi$ is a correlation length, $\phi_0$ is the flux quantum,  and $\mu$ is the vortex mobility.  $n_f$ is the areal density of thermally excited free vortices, which is related to the correlation length by the relation $n_f =1 / 2\pi c_1 \xi^{2}$.  $c_1$ and $c_2$ are small dimensionless constants.  It is reasonable to expect that very close to $T_c$ vortices are the principal degrees of freedom in even quasi-2D materials.  Note that these are essentially model-free forms constrained only by dimensional analysis, Maxwell equations, and immutable properties of superfluid vortices like the Josephson relation.  Using accepted values for $c_1$ and $c_2$ \cite{Halperin79a}, and the Einstein relation $D = \mu k_B T$, the expression 
 \begin{equation}
D(T) = -  \frac{6}{ \mu_0 }  \frac{\chi_{2D}}{G_S}.   
\label{ratio}
\end{equation}

 \noindent follows \cite{Orenstein06a} and in principle may be used to give a determination of the vortex diffusion constant $D$ using only experimentally determined quantities.  Interestingly, this treatment using the analogous equations within the Gaussian approximation and in the dirty limit gives the diffusion constant of the normal state $electrons$.  This is potentially useful as a diagnostic considering that electronic diffusion is proportional to the normal-state conductance while vortex diffusion is conventionally proportional to the normal-state resistance.  One may also heuristically motivate Eq. \ref{ratio} through the fact that correlations in length (diamagnetism in 2D $\propto \xi^2$) probed by a thermodynamic measurement like susceptibility and the correlations in time ($1/\Omega$) probed by a dynamic measurement like conductivity are related within diffusive dynamics as $\xi^2 \propto D/ \Omega$, where $\Omega$ is the characteristic fluctuation rate. 
 
A problem with applying Eq. \ref{ratio} to real type-II superconductors is that, in general, the motion of vortices is limited by both dissipative (viscous) flux-flow and pinning forces.  In 2D, the classical equation of motion for a single vortex is $ \dot{x} / \mu + k_p x = K_y \phi_0$ where $K_y$ is a driving sheet current, $x$ is the vortex displacement and $k_p$ is a pinning constant \cite{Gittleman66a}.  Here the complex physics of pinning and flux-flow are represented by phenomenological parameters.  This leads to an expression for the 2D resistance from moving vortices as $R_v = \phi_0^2 n_f  \mu [ 1 / (1+ i \omega_{d}/\omega)]$ where $\omega_{d} = k_p \mu $ is the ``depinning frequency".   This expression shows that at frequencies well above $\omega_{d}$, viscous forces dominate and the motion of vortices becomes predominately dissipative.   This is a considerable simplification.  In this limit the expression for $R_v$ reduces to the inverse of Eq. 2 for the vortex conductance.  In cuprate superconductors, $\omega_{d}$ is generally of the order of a few GHz \cite{Golosovsky94a}.  This puts the appropriate frequency regime to probe purely dissipative vortex transport in the range of our TTDS measurements.

We have measured the THz range optical conductivity of molecular beam epitaxy (MBE) grown LSCO films using a homebuilt transmission-based time-domain THz spectrometer.  With this technique the complex transmission function can be directly inverted to get the complex conductivity \cite{Comment2}.  In Fig. 1(a) and (b) we present the real ($\sigma_1$) and imaginary ($\sigma_2$) THz conductivity of one particular LSCO film (x=0.095, $T_c$=23.5K) out of a large series we have recently studied \cite{Bilbro11a}.  At high temperature $\sigma_1$ is fairly constant in frequency.  As the temperature is lowered, $\sigma_1$ increases, develops a frequency dependence near $T_c$, and then decreases as spectral weight is shifted into a delta function at zero frequency.  The $\sigma_2$ vs. frequency data in Fig. 1(b) show a small imaginary part of the conductivity at high temperatures, which is enhanced dramatically as temperature is reduced near $T_c$.  At the lowest displayed temperatures $\sigma_2$ shows the 1/$\omega$ dependence expected for the superfluid response of a superconductor.  While the low and high-temperature limits are easily understood, we are most interested in the fluctuation regime near $T_c$.

The enhancement of the conductivity in this fluctuation regime is more clear in Fig. 1(c) and (d), where we plot $\sigma_1$ and $\sigma_2$ vs. temperature.  One can see clearly the slow increase and subsequent decrease in $\sigma_1$ as temperature is lowered below $T_c$.  At low frequency there is a well-defined peak around $T_c$.  The location of this peak shifts to lower temperature as frequency is reduced corresponding to the slowing down of fluctuations as the temperature decreases.  Above $T_c$, we see a sudden onset in $\sigma_2$ at a temperature $T$ $\approx$ 30K.  In earlier work, we found that the second derivative with respect to temperature of the quantity $\omega\sigma_2$ (which is related to the phase stiffness) showed a clear and dramatic onset from a near-zero high-temperature signal \cite {Bilbro11a}.  We denoted this temperature as $T_o$, and defined it as the onset of superconducting fluctuations in the charge conductivity (for this film $T_o\approx$ 31K).  Note that there is no sign of conductivity enhancement at the high temperatures of the Nernst or diamagnetism onset \cite {Xu00a,Wang05a,Li10a}.

\begin{figure}[t]
\includegraphics[width=1\columnwidth]{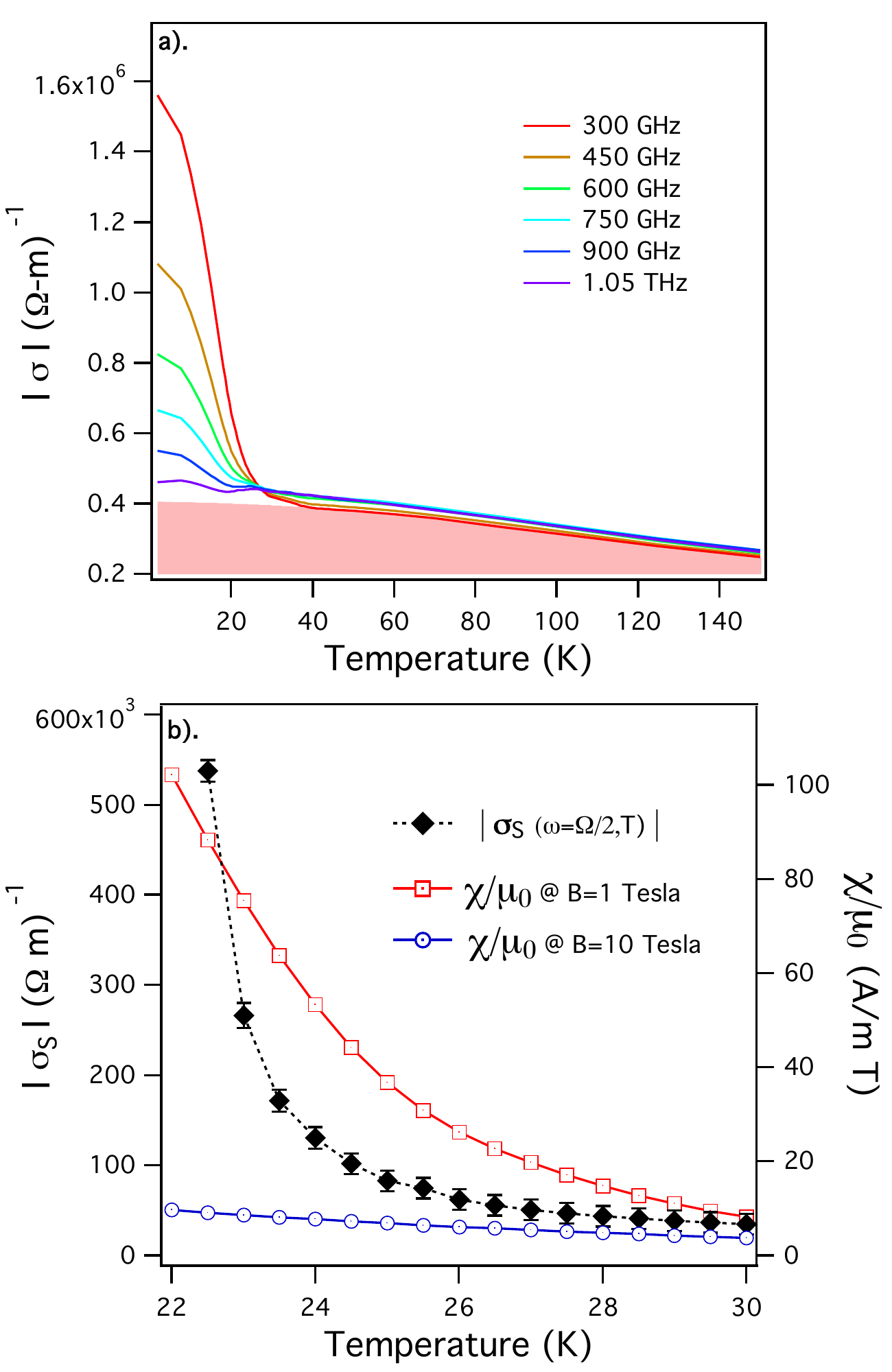}
\caption{(a) Magnitude of the conductivity ($|\sigma|$) as function of temperature.   The filled region is a fit of the normal state background conductivity at 300 GHz \cite{Comment2}.  The fluctuation conductivity $\sigma_S$ is obtained by subtracting this background from $|\sigma|$.  (b) A comparison of fluctuation conductivity with the diamagnetism in similarly doped La$_{2-x}$Sr$_x$CuO$_4$ crystals \cite{Ong10a}.}
\end{figure}

As mentioned above, in conventional models where vortices are the principal degree of freedom in the region above $T_c$, one expects that correlations in length and time scale together as a diffusionlike relation with vortex diffusion constant $D$.  Therefore, the large difference in the temperature of the inferred onset of superconducting correlations $T_o$ above $T_c$ between our experiments (10 - 20K) and for instance, diamagnetism measurements ($\approx$100K)  \cite{Wang05a,Li10a} begs an explanation.   Here we evaluate the relative size of the signals in terms of the diffusion constant derived in the above analysis and show that conventional vortex dynamics would predict a much larger fluctuation conductivity given the size of diamagnetism.

\begin{figure}[t]
\begin{center}
\includegraphics[width=1\columnwidth]{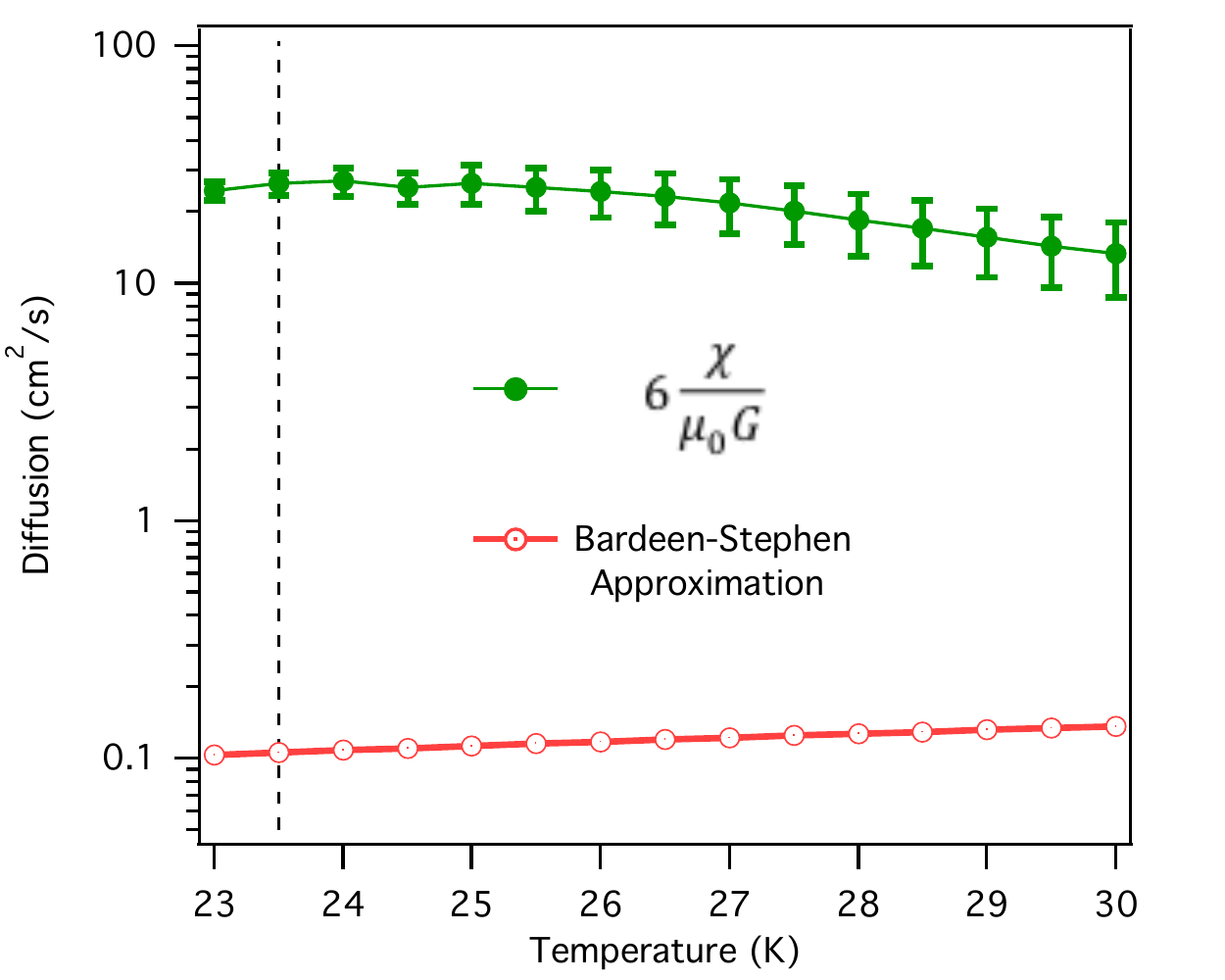}
\caption{Comparison of the Bardeen-Stephen estimation and calculated diffusion constant using our measured fluctuation conductivity, the estimated normal state background from the OSI and the 1 Tesla magnetic susceptibility.}
\end{center}
\end{figure}

In Fig. 2(a), we plot the magnitude of the conductivity $|\sigma|$.   To isolate the superconducting fluctuation contribution $\sigma_S$, we define a normal-state contribution that fits the conductivity well at temperatures above the onset of the diamagnetism (T$_D \approx$ 75 - 110 K in this doping range \cite{Li10a}), extrapolate to low temperatures, and take the difference.   Although we fit the background through a temperature-dependent Drude model \cite{Comment2}, our final conclusions are not sensitive to the precise background choice as we are only concerned with the temperature region up to about 10 K above $T_c$, where the fluctuations are obvious.

In previous work \cite{Bilbro11a} we have performed a scaling analysis that allowed us to extract the characteristic frequency scale $\Omega$ of the fluctuation superconductivity in the region above T$_c$ \cite{Comment2}.  In the analysis that follows we evaluate $|\sigma_S (\omega,T)|$ at a frequency $\omega=\Omega(T)/2$ for each temperature.  This conductivity differs formally from the conductivity in Eq. \ref{Cond}  by a constant of order unity, which we set to one below.  The use of THz frequencies eliminates the effects of pinning and the scaling analysis essentially connects the response of the system at finite frequency to the dc response that the system $would$ have had in the absence of vortex pinning.  In Fig. 2(b), on the left axis, we plot the magnitude of the fluctuation conductivity contribution, evaluated at $\omega=\Omega(T) / 2$, vs. temperature.  On the right axis, we include diamagnetic susceptibility $\chi/\mu_0$ at 1 and 10 Tesla of a single-crystal LSCO sample \cite{Ong10a} with a similar doping and $T_c$ ($x=0.9$ and 23 K, respectively).  In this data, one can see how the larger field suppresses the susceptibility near $T_c$.  Although there is some correspondence between the form of the lower-field susceptibility with the conductivity, we now show that in fact it is the relative scale of these quantities which is particularly remarkable.

We now apply Eq. 3 with the data in Fig. 2(b) to extract $D$ for a small range of temperatures above $T_c$.  As shown in Fig. 3 we find that $D$ is of the order of 10's of cm$^2$/sec throughout the range above $T_c$.  This is at least 2 orders of magnitude larger than a simple Bardeen-Stephen (BS) estimate $D =  (2 k_B T e^2 \xi^2_{c}) / (\pi \hbar^2  \sigma_n t)$ \cite{Stephen65a} (here $\sigma_n$ is the extrapolated normal state background conductivity, $t$ is the spacing between CuO$_2$ layers, and $\xi_{c} $ is the vortex core size \cite{Kato04a, Pan00a}).  The BS approximation appears to work well to model flux-flow dissipation in conventional $s$-wave materials \cite{Peroz05a,Fiory68a, Berghuis93a}, where the majority of dissipation occurs through quasiparticle motion in the vicinity of the vortex cores.  Note that the magnetic susceptibility appears to become singular as $B \rightarrow$ 0 near T$_c$ (``fragile London rigidity") \cite{Li10a}, so that evaluating $D$ at lower fields (corresponding to our $B=0$ TTDS experiment) will only increase the ratio of $\chi/G$ and the discrepancy with the BS estimate.  Although there is an expectation that due to their $d$-wave nature, short coherence lengths, gapped vortex core, and proximity to the Mott insulator, the cuprate vortices may be ``fast" as compared to the BS estimate \cite{Volovik93a, Maggio95a, Ioffe02a, Melikyan05a, Lee06a, Nikolic06a, Fanfarillo2011a}, the discrepancy we find is extreme.  It is an open question whether a diffusion constant as large as we have found can be reconciled.  We have currently performed this analysis for one underdoped sample due to difficulty in obtaining compatible diamagnetism data.  However, we anticipate similar behavior for the entire underdoped part of the phase diagram, since signals of conductivity and diamagnetism vary smoothly as a function of doping \cite{Li10a,Bilbro11a}.

There are two obvious possible conclusions from our data and analysis.   If in fact the large diamagnetic response in the cuprates comes entirely from superconducting correlations, then we have shown that their vortex motion must be anomalously fast and their dissipation anomalously small to reconcile the behavior with charge transport.  One expects that above $T_c$ an effective two fluid model may apply where the total conductivity has contributions from both normal electron and superconducting degrees of freedom in the form of $\sigma_T = \sigma_N + G_S / t$ where $G_S$ is given by Eq. 2 in the vortex regime.   Our results show the manner in which superconducting correlations may persist far above $T_c$ but be invisible to the charge response; the fast vortices are shorted out by the normal electrons.  It is a separate, but well posed theoretical challenge to explain vortex motion this fast.  Although detailed calculations must be performed, it is possible that such anomalously fast diffusion may arise as a consequence of the cuprates' $d$-wave nature and small gapped cores \cite {Melikyan05a, Geshkenbein98a}, inhomogeneities \cite{Martin10a}, the existence of a competing state nucleated in the vicinity of a vortex \cite{Lee06a}, or the proximity to a Mott insulator \cite{Ioffe02a, Fanfarillo2011a}.  Alternatively, if calculations show that vortex dissipation must always be at least parametrically related to the BS estimate by numbers of order unity, then our analysis shows that there must be another large contribution to the diamagnetic response that is not superconducting in origin (See Ref. \cite{Sau2010a} for one such possibility).

We thank L. Li, I. Martin, A. Millis, P. Nikolic, V. Oganesyan, N.P. Ong, O. Pelleg, Z. Tesanovic, and S. Tewari for helpful discussions and/or correspondences.  We also would like to thank L. Li and N.P. Ong for access to their unpublished data.  Support for the measurements at JHU was provided by 10DOE DE-FG02-08ER46544 and the Gordon and Betty Moore Foundation.   The work at BNL was supported by U.S. DOE under Project No. MA-509-MACA.

\clearpage

\subsection{Online supplementary information for ``A vortex plasma model analysis of the THz conductivity and diamagnetism in  La$_{2-x}$Sr$_{x}$CuO$_4$" }

\paragraph{Time-domain THz spectroscopy - }

We have measured the THz range optical conductivity using a home-built transmission based time-domain THz spectrometer.   In this technique, we split a femtosecond laser pulse along two paths and sequentially excite a pair of photoconductive `Auston'-switch antennae on radiation damaged silicon on sapphire.  A broadband THz range pulse is emitted by one antenna, transmitted through the LSCO film, and measured at the other antenna.  By varying the length-difference of the two paths, the electric field of the transmitted pulse is measured as a function of time.  Ratioing the Fourier transform of the transmission through the LSCO film on a substrate to that of a bare reference substrate, we resolve the frequency dependent complex transmission.  The transmission is inverted to obtain the complex conductivity by the standard formula for thin films on a substrate: $\tilde{T}(\omega)=[(1+n)/(1+n+Z_0\tilde{\sigma}(\omega)d)] e^{i\Phi_s}$ where $\Phi_s$ is the phase accumulated from the small difference in thickness between the sample and reference substrates and $n$ is the substrate index of refraction.

\paragraph{Scaling analysis of the fluctuation conductivity - }

In previous work \cite{Bilbro11aSI} we have also performed a scaling analysis that allowed us to extract out a characteristic fluctuation rate of the superconductivity.   This analysis follows from the fact that for a fluctuating superconductor one expects that the relation
\begin{equation}
\sigma_S(\omega)= \frac{G_Q}{t} \frac{k_B T_\theta^0}{\hbar \Omega} \mathcal S (\frac{\omega}{\Omega})
\label{scaling}
\end{equation}

\noindent holds for the portion of the conductivity, $\sigma_S$, due to superconducting fluctuations.  Here $G_Q = e^2/ \hbar$ is the quantum of conductance, $t$ is the inter-CuO$_2$ plane spacing, $T_\phi^0$ is a temperature dependent prefactor and  $\Omega$ is the characteristic fluctuation rate.  This scaling function is similar to the one proposed by Fisher, Fisher, and Huse \cite{Fisher91aSI} and is identical to the one used in previous THz measurements on underdoped BSCCO \cite{Corson99aSI}.  In Fig. 4a we show the collapsed phase $\varphi =$ tan$^{-1}\sigma_2/\sigma_1$ from the data in Fig. 1 as a function of reduced frequency $\omega/\Omega$ at temperatures from 22 K to 30 K.  The phase is an increasing function of $\omega/\Omega$, with the metallic limit $\varphi = 0$ reached at $\omega/\Omega \rightarrow 0$ and $\varphi$ becoming large (but bounded by $\pi/2$) as  $\omega/\Omega \rightarrow \infty$.  We plot the extracted fluctuation rate $\Omega$ as a function of temperature in Fig. 4b.  As noted previously \cite{Bilbro11aSI}, we continue to obtain good scaling and data collapse if we push the analysis 3-4 K above the temperature of the obvious onset in $\sigma_2$.   This region shows a linear dependence of $\Omega$ on temperature.

\begin{figure}[h]
\begin{center}
\includegraphics[width=1\columnwidth]{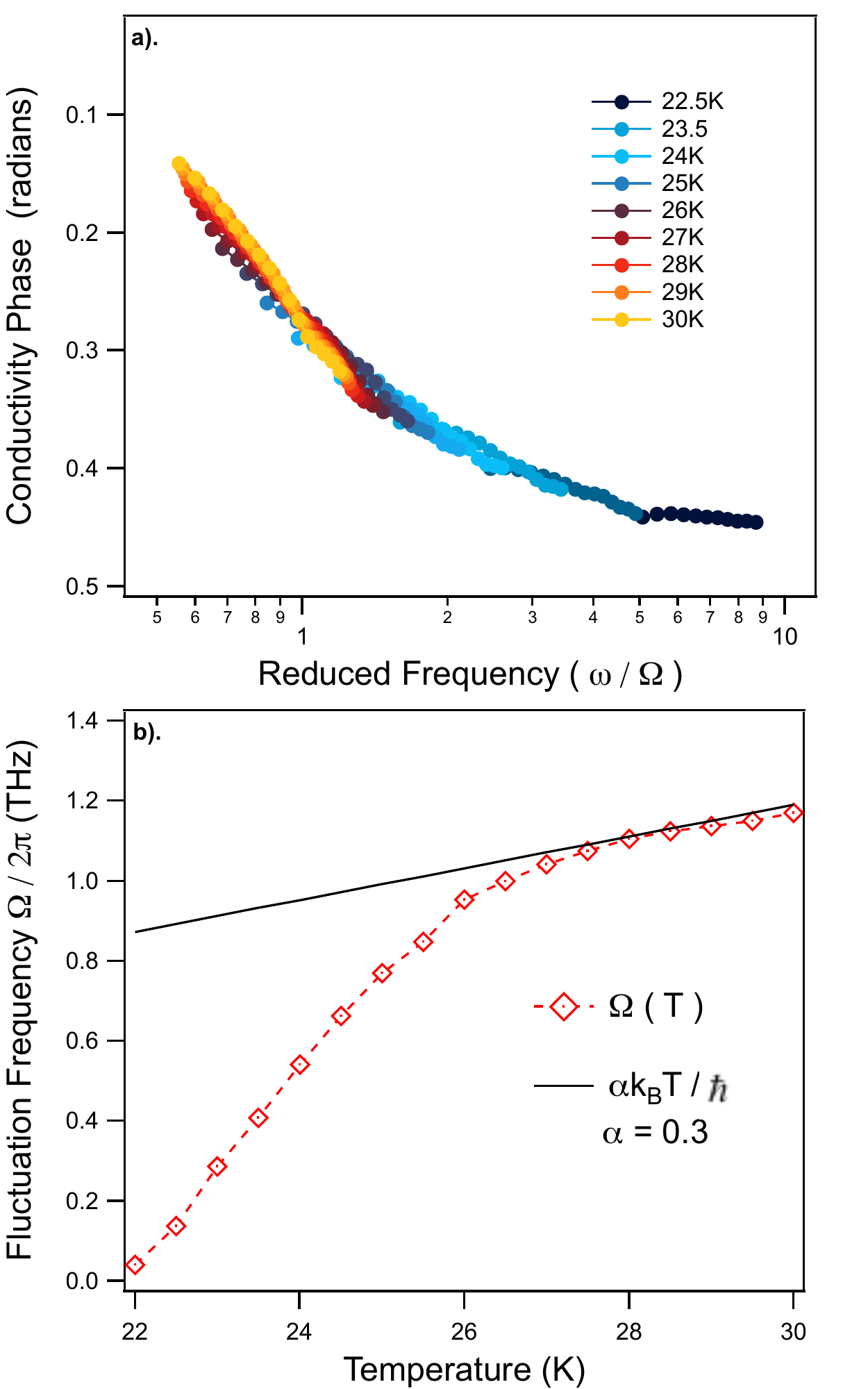}
\caption{(a) The conductivity phase $\varphi =$ tan$^{-1}\sigma_2/\sigma_1$  vs. scaled frequency $\omega/\Omega$ as described in the text. (b) The extracted fluctuation rate $\Omega$ obtained from the proportionality of the collapsed phase and the scaling function S($\omega/\Omega$).}  \end{center}
\end{figure}

In the main text, we use $\Omega(T)$ to evaluate the magnitude of the fluctuation contribution to the conductivity.   We evaluate the magnitude at a frequency of fixed proportionality of 1/2 the characteristic fluctuation frequency at each temperature, i.e. $\sigma_S(\omega=\Omega(T)/2 ,T)$.  The $S$ function in Eq. 4 then becomes the same constant of order unity for all temperatures;  in the analysis in this paper we set it equal to one.

\paragraph{Modeling of normal state conductivity - }

We estimate the normal state contribution to the conductivity by using the Drude model $\sigma = \frac{\omega_p^2}{4 \pi} \frac{\tau}{1 - i \omega \tau}$, which should be valid in the cuprate normal state at low enough frequencies.  In this model note that $\sigma_2(\omega)/\sigma_1(\omega) = \omega\tau$.  Assuming $\tau$ is relatively frequency independent in this frequency range, we extract $\tau$ at high temperatures by finding the slope of $\frac{\sigma_2}{\sigma_1}(\omega)$.  Using the high temperature $\tau$ and $|\sigma|$, we find the plasma frequency $\omega_p$.  We then fit $1/\tau$ to a power law form a$_1$+b$_1$T$^n$ and $\omega_p$ to a linear form a$_2$+b$_2$T at high temperature and extrapolate these fits to low temperature.  Since $\omega\tau << 1$ in this frequency regime, the normal state background contribution is well approximated as $|\sigma_{bg}| = \omega_p^2\tau/4\pi$.  The error bars on the vortex diffusion and fluctuation conductivity come from our uncertainty in fitting $1/\tau$ and $\omega_p$ in the high temperature range.  The upper limit of $|\sigma_{bg}|$ was set with the values of the $1/\tau$ fitting parameters (in THz units) of a$_1$=4.08, b$_1$=1.33$\times 10^{-6}$, and n=2.82; the lower limit of $|\sigma_{bg}|$ used the values a$_1$=3.72, b$_1$=2$\times 10^{-4}$, and n=1.85.  In fitting $\omega_p$ we allowed a$_2$ to have a very small frequency dependence (varying by $\approx 2\%$ of the average value of 4.5 GHz), while b$_2$ was kept constant at b$_2$=-1.09 MHz/K.

\paragraph{Molecular beam epitaxy of La$_{1.905}$Sr$_{0.095}$CuO$_4$ films - }

The LSCO films were deposited on 1-mm-thick single-crystal LaSrAlO$_4$ substrates, epitaxially polished perpendicular to the (001) direction, by atomic-layer-by-layer molecular-beam-epitaxy (ALL-MBE) \cite{Bozovic01aSI}.  The samples were characterized by reflection high-energy electron diffraction, atomic force microscopy, X-ray diffraction, and resistivity and magnetization measurements, all of which indicate excellent film quality. The thickness is known accurately by counting atomic layers and RHEED oscillations, as well as from so-called Kiessig fringes in small-angle X-ray reflectance and from finite thickness oscillations observed in XRD pattern.  For further details see Refs. \cite{Bilbro11aSI,Bozovic01aSI}.

\end{document}